\documentclass{cambridge6A}
\usepackage{graphics}
\usepackage[active]{srcltx}
\usepackage{amsfonts}
\usepackage{amsmath}
\usepackage{bm}
\input{epsf}
\usepackage{epsfig,here}

\begin{document}

\chapter{Protective Measurement, Postseletion and the Heisenberg Representation}

\begin{center} \large Yakir Aharonov$^{1,2}$, Eliahu Cohen$^1$ \end{center}

 \begin{center} $^1$School of Physics and Astronomy, Tel Aviv University, Tel Aviv 69978, Israel, eliahuco@post.tau.ac.il \\
                $^2$Schmid College of Science, Chapman University, Orange, CA 92866, USA, yakir@post.tau.ac.il \end{center}

Classical ergodicity retains its meaning in the quantum realm when the employed measurement is protective.
This unique measuring technique is reexamined in the case of post-selection, giving rise to novel insights studied in the Heisenberg representation.
Quantum statistical mechanics is then briefly described in terms of two-state density operators.

\section{Introduction}

In classical statistical mechanics, the ergodic hypothesis allows us to measure position probabilities in two equivalent ways: we can either
measure the appropriate particle density in the region of interest or track a single particle over a long time and calculate the proportion of
time it spent there. As will be shown below, certain quantum systems also obey the ergodic hypothesis when protectively measured. Yet, since
Schr\"{o}dinger's wavefunction seems static in this case \cite{prot1, prot2, prot3}, and Bohmian trajectories were proven inappropriate for calculating time averages of
the particle's position \cite{bohm1, bohm2}, we will perform our analysis in the Heisenberg representation.

Indeed, quantum theory has developed along two parallel routes, namely the Schr\"{o}dinger and Heisenberg representations, later shown to be equivalent.
The Schr\"{o}dinger representation, due to its mathematical simplicity,  has become more common. Yet, the Heisenberg representation offers some
important insights which emerge in a more natural way, especially when employing modular variables \cite{modular}. For example, in the context
of the two-slit experiment it sheds a new light on the question of momentum exchange \cite{Scully,Durr,Herzog}. Recently studied within the
Heisenberg representation are also the Double Mach-Zehnder Interferometer \cite{DMZI} and the N-slit problem \cite{A-tonomu}.
As can be concluded from \cite{A-tonomu}, the Heisenberg representation prevails in emphasizing the nonlocality in quantum mechanics thus providing us with
insights about this aspect of quantum mechanics as well.

Equipped with the backward evolving state-vector within the framework of Two-State-Vector Formalism (TSVF) \cite{TSVF}, the Heisenberg representation
becomes even more powerful since the time evolution of the operators includes now information from the two boundary conditions. Furthermore, when performing
post-selection, deeper understanding of the quantum system becomes available, such as the past of a quantum particle \cite{Vaidman1,Vaidman2}.

Post-selection does not change the protective measurement's results, but suggests interpreting them differently, thus enabling us
to effectively sketch two wavefunctions rather than one in the Schr\"{o}dinger representation. In the Heisenberg representation, a full description of time-dependent operators
emerges which enables further insights. Choosing a specific final state amounts to outlining another (sometimes, completely different) history
for the same initial state, that is, a different set of characterizing weak values.
In what follows, we use the Heisenberg representation to study protective measurements with post-selection. This way, we regain quantum
ergodicity and describe two-state ensembles coupled to a heat bath.

The rest of the paper is organized as follows: Sec. 2 discusses the differences between classical and quantum ergodicity. Sec. 3 describes
protective measurement in the Heisenberg representation. Cases of post-selection and external protection are analyzed. In Sec. 4 we show how to
describe quantum statistical mechanics in terms of two-state vectors. Protective measurement is utilized for studying the two-state density
operator and the resulting ensemble averages. Sec. 5 summarizes the main contributions of this work into a coherent description of protective measurement
in the Heisenberg representation.

\section{Classical and Quantum Ergodicity}

 We begin by examining a classical gas, i.e. an ensemble of $N$ point-like particles. Each individual particle is characterized
 by its position and momentum, so that in each moment the system can be described by a point in the $6N$-dimensional phase-space. The time
 average of a certain property $A$ over a time interval of length $T$ is given by:

\begin{equation}\label{time}
\bar{A} = \lim_{n\to\infty}\frac{1}{n} \sum_{j=1}^{n}A(\frac{jT}{n})
\end{equation}

Therefore, in order to accurately find $\bar{A}$ we ought to perform a large number of $A$ measurements at different times.

Under the ergodic assumption \cite{ergodic} this average is equivalent to the ensemble average at a certain moment:

\begin{equation}\label{ensemble1}
\langle A \rangle = \sum_{j=1}^{N}\frac{A_j}{N}
\end{equation}

More generally,

\begin{equation}\label{ensemble2}
\langle A \rangle = \int A d\mu,
\end{equation}

where $\mu$ is some finite, non-zero probability measure.

Is this reasoning applicable also in the quantum realm? First, in order to incorporate uncertainty, the phase-space should be partitioned
into hypercubes of volume $\hbar^{6N}$. Second, a practical question has to be addressed: how to perform all the measurements needed for an
accurate time average on a single particle without disturbing it?
This is where a resolution can be achieved with the help of protective measurement suggested for the first time by Aharonov and Vaidman in 1993
\cite{prot1} and further developed in \cite{prot2, prot3, prot4, prot5}. Moreover, using protective measurement it was argued that the wavefunction
should be understood as describing the (discontinuous, random in nature) ergodic motion of a single particle \cite{Gao}.

\section{Protective Measurements in the Schr\"{o}dinger and Heisenberg representations}

Protection of the state in the case of discrete non-degenerate spectrum of energy eigenstates was shown to be a consequence of energy
conservation when the measurement is sufficiently slow and weak \cite{prot2}. Protection can be achieved also in more general cases by utilizing a
protective interaction term in the Hamiltonian.
This possibility of performing a dense set of measurements without affecting the measured state, allowed ``observing'' the wave function \cite{prot1}.
In the Schr\"{o}dinger representation it seems that the evolution of the wave function was tightly restricted, what let us later obtain its form
everywhere in space.
Putting it in more formal terms, protective measurement can be carried out by applying an interaction Hamiltonian of the form:

\begin{equation}\label{Hint}
H_{int} = g(t)pP_{V_i}
\end{equation}
with $g(t)=1/T$ for a period of $T$ smoothly approaching zero before and after the measurement. Where $p$ is the momentum of the measuring pointer,
$P_{V_i}$  is the projection operator into the set $V_i$, and $V=\sum_{n}V_i$ is the total space region.
Let us assume that the system in question is an harmonic oscillator, and the initial wavefunction is
the ground state $|\psi_{in}\rangle=|0\rangle$, i.e. $\psi(x)={\pi}^{-1/4}e^{-x^2/2}$ (throughout the calculations we used $\sqrt{\hbar/m\omega}=1$)

Suppose also that we are interested in some remote $V_i$ centered around $x_0 \gg 1$ i.e. far from the origin. The particle has a small
probability to be found in that place, but when the measurement is long enough, we would find that the state of the pointer propagated in time
according to:

\begin{equation}\label{pointer}
U=e^{-\frac{i}{\hbar}p\langle P_{V_i}\rangle},
\end{equation}
although the energy has only changed negligibly for each $p$:

\begin{equation}\label{deltaE}
\delta E=\langle H_{int}\rangle=\frac{\langle P_{V_i} \rangle p}{T}.
\end{equation}

This way we can gain knowledge of $|\psi_{in}\rangle$ of a single particle in $V_i$. Repeating this measurement for all $V_i$ we would finally
be able to sketch $|\psi_{in}\rangle$ in $V$.

Here we introduce post-selection in the form of slicing past events using a certain final state \cite{unified,future}. By this we mean grouping together
all the experiments which ended at the same state. What does it change? Clearly, the results of the protective measurement do not change, giving
rise to the same observation of the wave-function. The ontology however, turns out to be different.
Our initial state was: $|\psi_{in}\rangle=|0\rangle$. When performing the trivial post-selection, that is, $|\psi_{fin}\rangle=|\psi_{in}\rangle$, within
the Schr\"{o}dinger representation we believe that the protective measurement probed a static (up to a changing phase) eigenstate of the oscillator
having a small probability to be found in $V_i$. Hence, the pointer translation grew slowly but surely according to Eq. \ref{pointer}.
However, suppose we postselect a different final state which is some coherent state $|\alpha\rangle$ (since coherent states form an overcomplete basis,
this can be done approximately by defining the appropriate POVM). In our experiment, the final measurement will allow finding the initial state as a
coherent state $|\alpha\rangle$ with probability $e^{-|\alpha|^2/2}$. In the position representation, the coherent state is denoted at every moment by
\cite{Schwabl}:

\begin{equation}\label{coherent_x}
\varphi_{\alpha}(x,t)= {\pi}^{-1/4}exp\{-i\Theta(x,t)-\frac{1}{2}{[x-\sqrt{2}|\alpha|cos(\omega t-\delta)]}^2\}
\end{equation}
where $\alpha=|\alpha|e^{i\delta}$ and

\begin{equation}\label{Theta}
\Theta(x,t)= \frac{\omega t}{2} -\frac{{|\alpha|}^2}{2}sin[2(\omega t -\delta)]+\sqrt{2}|\alpha|xsin(\omega t -\delta)
\end{equation}

The same result of Eq. \ref{pointer} suggests now a significant motion along the harmonic well of this backward evolving coherent state.
As was shown in \cite{AAV, unusual}, any sufficiently weak coupling between a pointer and an observable $O$ of a pre- and post-selected quantum
system, is a coupling to a weak value:

\begin{equation}\label{weak_value}
O^w(t)=\frac{\langle\Phi_f(t)|O|\Phi_i(t)\rangle}{\langle\Phi_f|\Phi_i\rangle}
\end{equation}
where $|\Phi_i\rangle$ and $\langle\Phi_f|$ are the pre- and post-selected states respectively.
In order to demonstrate the movement of the pointer we shall assume its coupling to the real part of the weak value and find out:
\begin{equation}\label{weak_value}
Re\{P_{Vi}^w(t)\}=Re\{\frac{\langle\alpha ^* (t)|P_{Vi}|\psi_{in}(t)\rangle}{\langle\alpha ^*|\psi_{in}\rangle}\}
\end{equation}
that is:
\begin{equation}\label{coherent in Vi}
Re\{P_{Vi}^w(t)\} \approx \pi^{-1/2}e^{({|\alpha|}^2-{x_0}^2)/2}cos[\Xi(t)]e^{-{[x_0-\sqrt{2}|\alpha|cos(\omega(T-t)+\delta)]}^2/2}
\end{equation}
where
\begin{equation}\label{Xi}
\Xi(t)=\frac{\omega T}{2}+\frac{{|\alpha|}^2}{2}sin[2(\omega (T-t)+\delta)]-\sqrt{2}|\alpha|x_0sin[\omega(T-t)+\delta]
\end{equation}

\begin{figure}[h]
 \centering \includegraphics[height=6cm]{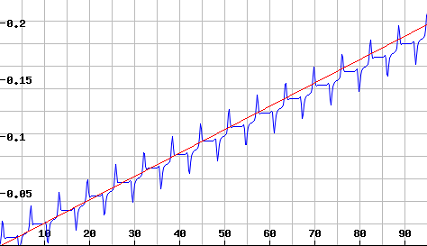}
      \caption{{\bf Pointer readings for two post-selections.} The pointer readings are shown for the trivial post-selection (red) and for the $\alpha$ post-selection (blue). Despite the different shape, they eventually reach approximately the same point.} \label{graphsA}
\end{figure}

Due to the oscillations of the post-selected coherent state, the pointer translation can be understood now to be nonlinear. According to Eq. \ref{coherent in Vi}
the pointer movement seems oscillatory (it moves each time the backward evolving coherent state ``pushes'' it), which is quite different form the
case of trivial post-selection where it moved linearly, so it finally reaches the same place as earlier, but with an altered history.
A comparison between the expectation value of the pointer readings in the case of trivial post-selection and in the case of $\alpha$ post-selection
is shown in Fig. \ref{graphsA}. For illustration purposes, the following parameters were chosen: $x_0=1$, $\alpha=2.5$, $\omega=1~Hz$ and $T=100~sec$.
We assume that the width of the pointer's wavefunction is large enough so that the measurement can be considered weak.

\begin{figure}[h]
 \centering \includegraphics[height=6cm]{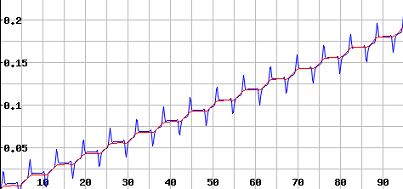}
      \caption{{\bf  A Comparison between $\alpha=2.5$ (blue) and $\alpha=1$ (red).}} \label{graphsB}
\end{figure}

\begin{figure}[h]
 \centering \includegraphics[height=6cm]{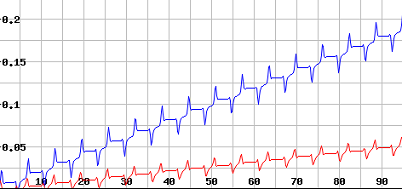}
      \caption{{\bf A Comparison between $x_0=1$ (blue) and $x_0=1.5$ (red).}} \label{graphsC}
\end{figure}

In order to better understand the movement which arises from Eq.\ref{coherent in Vi} we compare the results of the above $\alpha=2.5$ post-selection
to post-selection of $\alpha=1$ (while the other parameters remain the same). The forward and backward evolving states are now closer, so due to their
higher scalar product, the weak value, and hence the amplitude of oscillations, both decrease (see Fig. \ref{graphsB}).
Another comparison is drawn between the above case of searching for the wavefunction at $x_0=1$ to the case of searching at $x_0=1.5$. The chances to
find there the particle are now smaller, and therefore, the expectation value is lower (see. Fig. \ref{graphsC}).

Utilizing Bohr's correspondence principle, we could relate classical and quantum ergodicity: if instead of the ground state we would have
chosen a highly excited state (or alternatively, large $\alpha$ for the final state), we know, according to the correspondence principle, that the
classical time the oscillator spends in $V_i$ would be proportional to the relative number of harmonic oscillators, out of a large ensemble, that
could be found instantaneously within this interval.

This dynamic interpretation can be better understood within the Heisenberg representation.
First, we know that the operators $\hat{x}$ and $\hat{p}$ change in time just like the classical variables $x$ and $p$, hence ergodicity and correspondence
arise naturally.

Second, each projection operator $P_{V_i}(t)$ can be evaluated as a time-dependent matrix using the oscillator eigenstates:

\begin{equation}\label{Vi}
P_{V_i}^{m,n}(t)=e^{-i(m-n)t}\langle m| P_{V_i} |n \rangle.
\end{equation}
which in contrast to the evolution of the state seems very oscillatory. However, during the measurement interval, all the off-diagonal entries
tend to zero, and $P_{V_i}$ becomes approximately time-independent and diagonal. Therefore, after a long time its diagonal values directly
indicate ensemble averaging, thus expressing quantum ergodicity. This could also be understood from the coherent states evolution which covers
all phase space, thus allowing the operators in Heisenberg representation to take any possible value.
Slicing past results according to all the possible future results, divides the ensemble to several distinct sub-ensembles, each of which having
different weak value and hence, different history of the measuring pointer.

Another discrepancy between the two representations apparently arises in case the initial state is a superposition of different energy
eigenstates. Artificial Zeno-type protection is needed in the form of very frequent projective measurement on the state which will preserve it by
halting its evolution (the time scale of intervals between consecutive protections must be much smaller than the time scale of changing the
wavefunction due to its Hamiltonian). In the Schr\"{o}dinger representation, it seems that the state rarely changes due to this procedure, hence
protective measurements are performed again and again on one and the same static state. In contrast, calculation in the Heisenberg representation
describes the image of subsequent abrupt changes of the operator we wish to measure.

\section{Statistical Mechanics with Two-State-Vectors}

Assume now the system is coupled to a heat bath of temperature $T={(k\beta)}^{-1}$ and allowed to reach equilibrium. The system will be described by the Boltzmann thermal density operator:

\begin{equation}\label{StatDens}
\rho=\frac{e^{-\beta H}}{Tr(e^{-\beta H})}
\end{equation}
For the harmonic oscillator discussed above it equals \cite{THO}:
\begin{equation}\label{HStatDens}
\rho=(1-e^{-\beta\hbar\omega})\sum_{n=0}^{\infty}e^{-n\beta\hbar\omega}|n\rangle\langle n|
\end{equation}

If the measuring time is longer than the period of thermal fluctuations, the protective measurement will indicate the correct mixed state, that is, the pointer will move according to the thermal average of the measured quantity.
Alternatively, one can switch-off the coupling to the thermal bath before performing the measurement, and then the measurement will select a single pure state, rather than a mixture, according to the Boltzmann distribution.

Recalling the mapping between the averages calculated with this operator and the expectation values of the pure state \cite{THO}:

\begin{equation}\label{Psibeta}
|\psi_\beta\rangle={1-e^{-\beta\hbar\omega}}^{1/2}\sum_{n=0}^{\infty}e^{-n\beta\hbar\omega/2}|n\rangle
\end{equation}
we can perform protective measurements of this state and find out expectation values of thermal ensembles without disturbing them. A single
protective measurement was shown until now to describe the wavefunction of a single particle, and here it allows to acquire knowledge
about a large ensemble coupled to a heat bath.

What is the time-symmetric version of this density operator?
The TSVF \cite{TSVF} enables us to describe a quantum system in-between two strong measurements with the aid of weak measurements \cite{AAV}. It is a
symmetric formulation of quantum mechanics ascribing equal footing to the initial (forward evolving) and final (backward evolving) wavefunctions.
The two-state vector $\langle \Phi| \; |\Psi\rangle$ was shown in \cite{Reznik} to give rise to the density operator:

\begin{equation}\label{rho1}
\rho(t)=\frac{|\Psi(t)\rangle\langle\Phi(t)|}{\langle\Phi|\Psi\rangle},
\end{equation}
which evolves according to von Neumann equation just like the 1-state density operator:

\begin{equation}\label{dynamics}
i\hbar\frac{\partial\rho}{\partial t}=[H,\rho].
\end{equation}

In the double coordinate system it was shown to be:

\begin{equation}\label{rho2}
i\hbar\frac{\partial\rho(x',x'',t)}{\partial t}=[H(x',p')-H(x'',p'')]\rho(x',x'',t),
\end{equation}
where $\rho(x',x'',t)=\langle x'|\rho(t)|x''\rangle$.

The two-state density operator enables calculating weak values as follows:

\begin{equation}\label{Aw}
A_w=\frac{tr(A\rho)}{tr(\rho)}.
\end{equation}

Examining now a canonical ensemble with inverse temperature $T={(k\beta)}^{-1}$, the two-state density $\rho$ would take the form:

\begin{equation}\label{rho3}
\rho=\frac{exp\{-\beta [H(x',p')-H(x'',p'')]\}}{tr(exp\{-\beta [H(x',p')-H(x'',p'')]\})}.
\end{equation}
thus allowing us to calculate ensemble- and hence time- averages in the two-state Heisenberg representation when employing protective measurements.

\section{Discussion}

 The wavefunction as observed by protective measurements gains its meaning only when very long measurements or measurements over a large ensemble are
 performed. It is not possible to measure instantaneously the wavefunction of a single particle. This suggests that the wavefunction has either a
 statistic or an ergodic meaning. However, operators in the Heisenberg representation, do allow a description of single quantum particle at a single
 time. In addition, when pre- and post-selection are performed, the measuring pointer describes a distinct history of the system, depending on both
 backward and forward evolving wavefunctions. Furthermore, a single protective measurement allows to find the thermal state of an ensemble coupled to
  a heat bath, which leads to a full description of two-state thermal ensembles.

\hfill

{\bf Acknowledgements}
\\ We wish Avshalom C. Elitzur, Tomer Landsberger and Daniel Rohrlich for helpful comments and discussions. This work has been supported in part by the Israel Science Foundation Grant No. 1125/10.


\begin{thebibliography} {22}

\bibitem{prot1}
Y.~Aharonov and L.~Vaidman,
The Schrödinger Wave is Observable After All!
in {\it
Quantum Control and Measurement}, H. Ezawa and Y. Murayama (eds.) {\bf 99}, Elsevier Publ., Tokyo (1993).

\bibitem{prot2}
Y. Aharonov, J. Anandan, and L. Vaidman, Meaning of the Wave
Function,  Phys. Rev. A {\bf 47}, 4616 (1993).

\bibitem{prot3}
Y. Aharonov and L. Vaidman,
Measurement of the Schr\"{o}dinger Wave of a
Single Particle,
  Phys. Lett. A {\bf 178}, 38 (1993).

\bibitem{bohm1}
Y. Aharonov, M. O. Scully and B.G. Englert,
Protective Measurements and Bohm Trajectories
Phys. Lett. A {\bf 263}, 137 (1999).

\bibitem{bohm2}
Y. Aharonov, N. Erez N and M.O. Scully,
Time and Ensemble Averages in Bohmian Mechanics
Physica Scripta  {\bf 69}, 81-83 (2004).

\bibitem{modular}
Y. Aharonov, H. Pendleton and A. Petersen,
Modular Variables in Quantum Theory, Int. J. Th. Phys. {\bf 2}, 213 (1969).

\bibitem{Scully}
M.O Scully, B.G Englert and H. Walther,
Quantum Optical Tests of Complementarity, Nature {\bf 351}, (6322), 111-116 (1991).

\bibitem{Durr}
S. Durr, T. Nonn and G. Rempe,
Origin of Quantum-Mechanical Complementarity Probed by a ``Which-way'' Experiment in an Atom Interferometer, Nature, {\bf 395} (6697), 33-37 (1998).

\bibitem{Herzog}
T.J Herzog, P.G. Kwiat, H. Weinfurter and A. Zeilinger,
Complementarity and the Quantum Eraser, Phys. Rev. Lett. {\bf 75} (17), 3034-3037 (1995).

\bibitem{DMZI}
J. Tollaksen, Y. Aharonov, A. Casher, T. Kaufherr and S. Nussinov,   Quantum Interference Experiments, Modular Variables and Weak Measurements,
New J. Phys. {\bf 12}, 013023 (2010).

\bibitem{A-tonomu}
Y. Aharonov,
On the Aharonov-Bohm Effect and Why Heisenberg Captures Nonlocality Better Than Schr\"{o}dinger, in {\it Tonomura Memorial Book}, Eds. Y.A. Ono and K. Fujikawa (2013).

\bibitem{TSVF}
Y. Aharonov and L. Vaidman,
The Two-State Vector Formalism of Quantum Mechanics, in {\it Time in Quantum Mechanics}, J.G. Muga {\it et al.} eds., Springer, 369-412 (2002).

\bibitem{Vaidman1}
L. Vaidman, Past of a quantum particle, Phys. Rev. A {\bf 87}, 052104 (2013).


\bibitem{Vaidman2}
A. Danan, D. Farfurnik, S. Bar-Ad and L. Vaidman, Asking Photons Where Have They Been, Phys. Rev. Lett. {\bf 111}, 240402 (2013).

\bibitem{ergodic}
K.G. Kay, Toward a Comprehensive Semiclassical Ergodic Theory, J. Chem. Phys. {\bf 79}, 3026 (1983).

\bibitem{prot4}
Y. Aharonov, and L. Vaidman,
 Protective Measurements of Two-State
Vectors,    in {\it Potentiality,
Entanglement and Passion-at-a-Distance }, R.S.Cohen, M. Horne and J.
Stachel (eds.), BSPS 1-8, Kluwer  (1997).

\bibitem{prot5}
Y. Aharonov, J. Anandan, and L. Vaidman, The Meaning of Protective Measurements,
Found. Phys. {\bf 26}, 117-126 (1996).

\bibitem{Gao}
S. Gao, Meaning of the Wave Function, Int. J. Quantum Chem. {\bf 111}, 4124-4138 (2011).

\bibitem{unified}
E. Cohen and A.C. Elitzur, Strength in Weakness: Broadening the Scope of Weak Quantum Measurement, forthcoming, Phys. Rev. A.

\bibitem{future}
Y. Aharonov, E. Cohen, D. Grossman and A.C. Elitzur,
Can Weak Measurement Lend Empirical Support to Quantum Retrocausality
EPJ Web of Conferences {\bf 58}, 01015 (2013).

\bibitem{Schwabl}
F. Schwabl, Quantum Mechanics, 3rd edition, Springer, 54-56 (2012).

\bibitem{AAV}
Y. Aharonov , D. Albert and L. Vaidman, How the Result of a Measurement of a Component of a Spin 1/2 Particle Can Turn Out to Be 100?, Phys. Rev. Lett. {\bf 60}, 1351-1354 (1988).

\bibitem{unusual}
Y. Aharonov, E. Cohen and S. Ben-Moshe, Unusual Interactions of Pre- and Post-selected Particles, forthcoming, Proceedings of ICNFP2012 (2014).

\bibitem{THO}
 Y. Aharonov , E.C. Lerner, H.W. Huang and J.M. Knight, Oscillator Phase States, Thermal Equilibrium and Group Representations, J. Math. Physics {\bf 14}, 746-756 (1973).

\bibitem{Reznik}
B. Reznik and Y. Aharonov, Time-symmetric formulation of quantum mechanics, Phys. Rev. A {\bf 52}, 2538-2550 (1995).

\end{thebibliography}
\end{document}